\documentclass{article}
\usepackage{authblk}
\usepackage[utf8]{inputenc}
\usepackage{float}
\usepackage{graphicx} 
\usepackage{caption}
\usepackage{upgreek}

\title{Reflection imaging with a helium zone plate microscope}
\author[1]{Ranveig Flatabø}
\author[1]{Sabrina D. Eder}
\author[1]{Thomas Reisinger\thanks{Present affiliation: Institute for Quantum Materials and Technologies, Karlsruhe Institute of Technology, Eggenstein-Leopoldshafen, Germany }}
\author[2]{Gianangelo Bracco}  
\author[3]{Peter Baltzer}
\author[1]{Björn Samelin}
\author[1]{Bodil Holst \thanks{e-mail: Bodil.Holst@uib.no}}
\affil[1]{Department of Physics and Technology, University of Bergen, Bergen, Norway}
\affil[2]{Department of Physics, University of Genova, Genova, Italy}
\affil[3]{MB Scientific AB, Uppsala, Sweden}

\begin{document}
\date{}
\maketitle

\begin{abstract}
Neutral helium atom microscopy is a novel microscopy technique which offers strictly surface-sensitive, non-destructive imaging. Several experiments have been published in recent years where images are obtained by scanning a helium beam spot across a surface and recording the variation in scattered intensity at a fixed total scattering angle $\theta_{SD}$ and fixed incident angle $\theta_i$ relative to the overall surface normal. These experiments used a spot obtained by collimating the beam (referred to as helium pinhole microscopy). Alternatively, a beam spot can be created by focusing the beam with an atom optical element. However up till now imaging with a focused helium beam (referred to as helium zone plate microscopy) has only been demonstrated in transmission. Here we present the first reflection images obtained with a focused helium beam. Images are obtained with a spot size (FWHM) down to 4.7 $\upmu$m $\pm $ 0.5 $\upmu$m, and we demonstrate focusing down to a spot size of about 1 $\upmu$m. Furthermore, we present the first experiments measuring the scattering distribution from a focused helium beam spot. The experiments are done by varying the incoming beam angle $\theta_i$ while keeping the beam-detector angle $\theta_{SD}$ and the point where the beam spot hits the surface fixed - in essence, a microscopy scale realization of a standard helium atom scattering experiment. Our experiments are done using an electron bombardment detector with adjustable signal accumulation, developed particularly for helium microscopy.

\end{abstract}

\section{Introduction}
Neutral helium atom microscopy,  usually abbreviated SHeM for scanning helium atom microscopy \cite{barr2012desktop}, uses a low-energy beam ($<$ 0.1 eV) of ground state helium atoms to map the outermost electron density distribution of the sample surface \cite{holst2021material}. This makes it particularly suitable for imaging fragile and/or high-aspect ratio samples that are challenging to image with charged-particles scanning probes or light techniques.  For a recent review of SHeM see  \cite{palau2023neutral}.

The helium beam is generated by a free-jet expansion where a skimmer selects the central part of the expansion. Two different microscope configurations exist. In the simpler configuration, the helium pinhole microscope, the beam is collimated using a pinhole \cite{witham2011, barr2014design, fahy2015highly, prav123, barr2016}. The pinhole defines the spot size and the shape of the beam. Alternatively, in the zone plate helium microscope, the beam is focused on the sample. Due to the inherent properties of ground state helium (neutral, inert, weak polarization coefficient, and no permanent magnetic moment), it can at present only be focused effectively using a mirror \cite{holst1997atom} or by the means of diffraction \cite{doak1999towards}. It has been challenging to control the curvature of mirrors with high enough precision to achieve a tightly focused beam. Therefore, in the second configuration of the microscope, the beam is focused onto the sample surface using a so-called Fresnel zone plate \cite{doak1999towards,reisinger2010free}. A Fresnel zone plate is a circular diffraction grating with decreasing grating constant (zone width) for increasing radii. The first-order diffracted beam is used for imaging. The advantage of using a focused rather than a collimated beam is that higher intensity as well as higher resolution, higher depth of field, and better signal-to-noise ratio can be achieved~\cite{palau2016theoretical, palau2017theoretical, bergin2019method}.

SHeM images are formed by scanning the sample under the beam and recording the transmitted or reflected signal across a given angle. In reflection, sample topography is usually the main contrast mechanism \cite{eder2023sub}, assuming that multiple scattering does not occur. In such cases, the beam is mainly specularly scattered (incident beam angle with respect to the local surface normal, $\theta_i$, equals outgoing beam angle with respect to surface normal, $\theta_f$). When the scale of the surface roughness is larger than the lateral resolution of the instrument, the topography of the sample surface changes the direction of the specularly scattered beam (because the orientation of the local surface normal changes), generating an image from beam and detector occlusion, referred to in the literature as supra resolution  \cite{eder2023sub}. Alternatively, when the scale of the roughness is smaller than the lateral resolution, the very small wavelength (0.1 nm or less) leads to so-called sub-resolution contrast \cite{eder2023sub}, where Ångström variations in roughness cause the specuarly scattered beam to broaden enabling imaging of nanoscale topography \cite{eder2023sub, prav123}.  For crystalline samples, sub-resolution contrast can be obtained through diffraction \cite{bergin2020observation}. In high-aspect ratio samples multiple scattering can lead to shadowing effects \cite{witham2012increased,lambrick2021true,lambrick2020multiple}. 

The first images ever obtained with a SHeM were obtained using a zone plate microscope in transmission \cite{koch2008imaging}. In this work, we present the first reflection images from a zone plate helium microscope. Further, we show that our zone plate microscope is capable of generating spot sizes with FWHM down to about 1 $\upmu$m. Finally, we present the first experiments measuring the scattering distribution from a focused helium beam spot.  As a demonstration example, we show that the scattering distribution from a mechanically polished aluminium surface is radically different from that of a stainless steel foil.

\section{Experimental}

\label{exp}
\begin{figure}[H]
	\includegraphics[width=\textwidth]{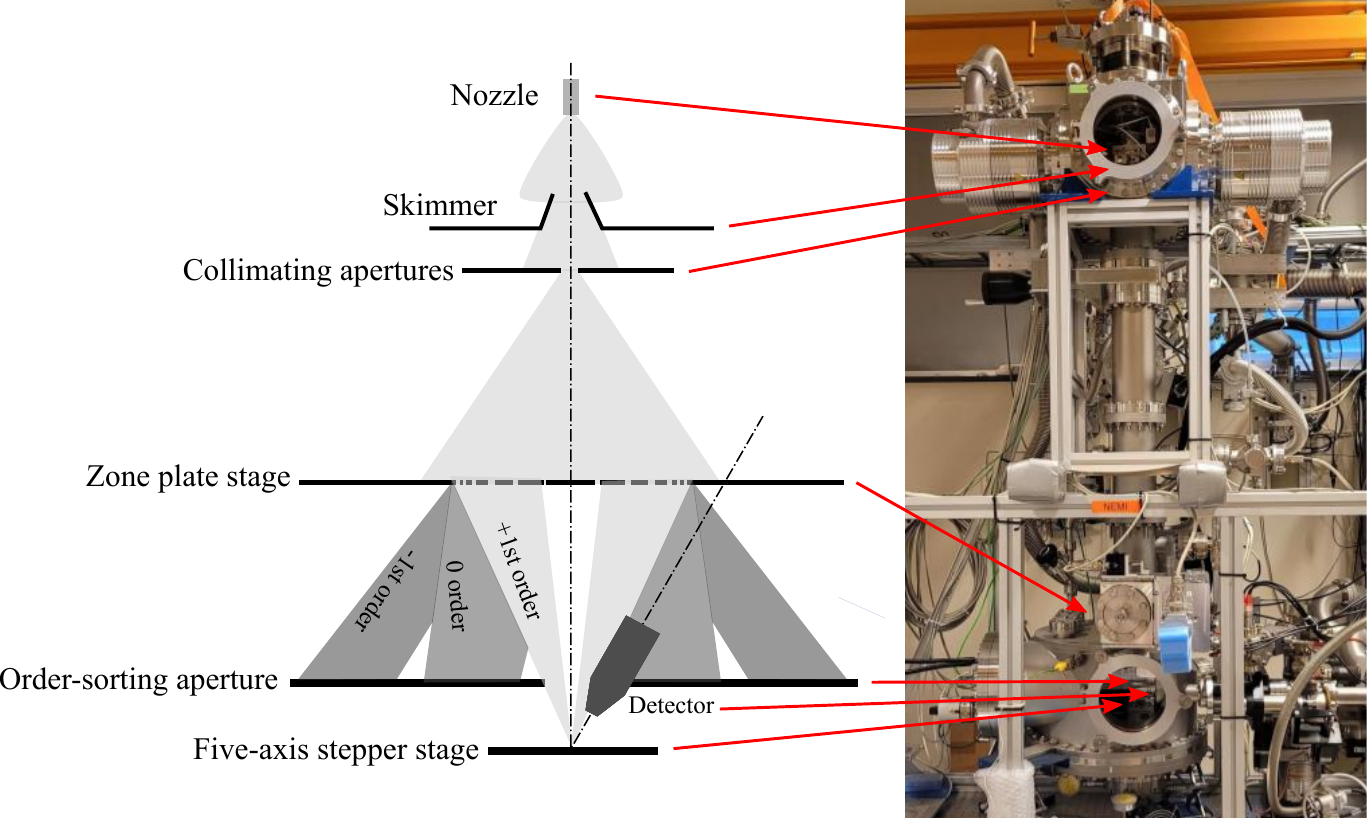}
	\caption{To the left, a schematic of the central components of the microscope. To the right a photo of the actual instrument. Each feature is highlighted in the image by the red arrows.   }
	\label{nemi}
\end{figure}

The main components of the zone plate SHeM are shown in Fig. \ref{nemi}. The neutral helium beam expands from a high-pressure gas reservoir through a 5 $\upmu$m $\pm$ 1 $\upmu$m nozzle into high vacuum (a free-jet expansion). For a detailed description of the source design, see~\cite{eder2013}. A skimmer (120 $\upmu$m in diameter) selects the central part of the expansion. The exchangeable collimating apertures range in size from 100 $\upmu$m to 5 $\upmu$m. They make it possible to change the spot size, and thereby the lateral resolution of the instrument, without breaking the vacuum, for a detailed description see \cite{flatabo2018fast}. The demagnification factor $M$ is defined as the ratio of the image distance (zone plate to sample distance) to the object distance (collimating aperture to zone plate distance). In the current configuration, the object distance is 907 mm and the image distance 205 mm, giving a demagnification factor of $M$ = 0.226.

The sample stage is a four-axis stepper stage from the company PI, which allows for linear movement in room coordinates X, Y, and Z. Additionally, a rotation axis is mounted on the Z axis for rotational movement. The incoming beam is aligned to be parallel to the stage Z axis, i.e., perpendicular to the sample holder.  Rotation movements move the sample holder away from the initial position (i.e. when rotation is zero) for axis X and Z. This movement is compensated for, making it possible to measure in one point while rotating the sample stage.

Scans are formed by rastering the sample under the beam and collecting the scattered signal. The reflection detector is custom-made and developed for this specific application by the company MB Scientific AB. It is located at a 30-degree angle relative to the incoming beam, and the scattered beam enters the detector via a 1 mm hole in a cone. The solid angle is 0.014 sr. The detector is an electron bombardment detector. It has two measuring modes, accumulating or not accumulating the signal, which can be selected depending on the incoming helium atom flux. The accumulation mode is used to increase the detector efficiency by a factor of 10 by increasing the time each helium atoms are exposed to electron bombardment. In non-accumulation mode, the total imaging time per image point including stage movements and settling time for detection is about 3 seconds. In accumulation mode, there is an additional 5-second wait time for the detected signal to be in equilibrium.

The fundamental resolution limit of the microscope is determined by the width of the outermost zone of the zone plate  \cite{michette1884optical}. A zone plate is designed for a given focal length and beam wavelength. The wavelength of the beam can be changed by changing the temperature of the source. The colder the source, the longer the wavelength (e.g. $\lambda$ = 0.055 nm at 320 K versus $\lambda$ = 0.098 nm at 100 K). A cold source is more intense than a warm source \cite{palau2018center}.

In this work, we use source temperatures of $\approx$ 320 K and $\approx$ 140 K. The warm (320 K) zone plate is designed with an outermost zone width of 48 nm and a diameter of 192 $\upmu$m. The cold zone plate is designed for an operating temperature of 100 K and has an outermost zone width of 125 nm and a diameter of 192 $\upmu$m \cite{reisinger2010free}. Both zone plates have a central middle-stop, 20~$\upmu$m in diameter, which blocks part of the 0-order contribution to the diffracted beam. This middle-stop is aligned along the optical axis (beam axis) with a 20~$\upmu$m diameter aperture, which forms the entrance to the sample chamber 162 mm down the beam axis from the zone plate (see Fig.~\ref{nemi}). This so-called order-sorting aperture ensures that the rest of the 0-order contribution as well as most of the higher-diffraction order contributions do not enter the sample chamber, thus reducing the helium background in the sample chamber and improving the signal-to-noise ratio. For a detailed description of the design see~\cite{eder2017}.  The source is cooled with liquid nitrogen. For the experiments presented here, we could only reach a temperature of about 140 K, which affects the focus and intensity. 

\section{Results and Discussion}

\subsection{Evaluating the spot sizes: Reflection image of a 10 $\upmu$m slit}

\begin{figure}[H]
\centering
\includegraphics[width=\textwidth]{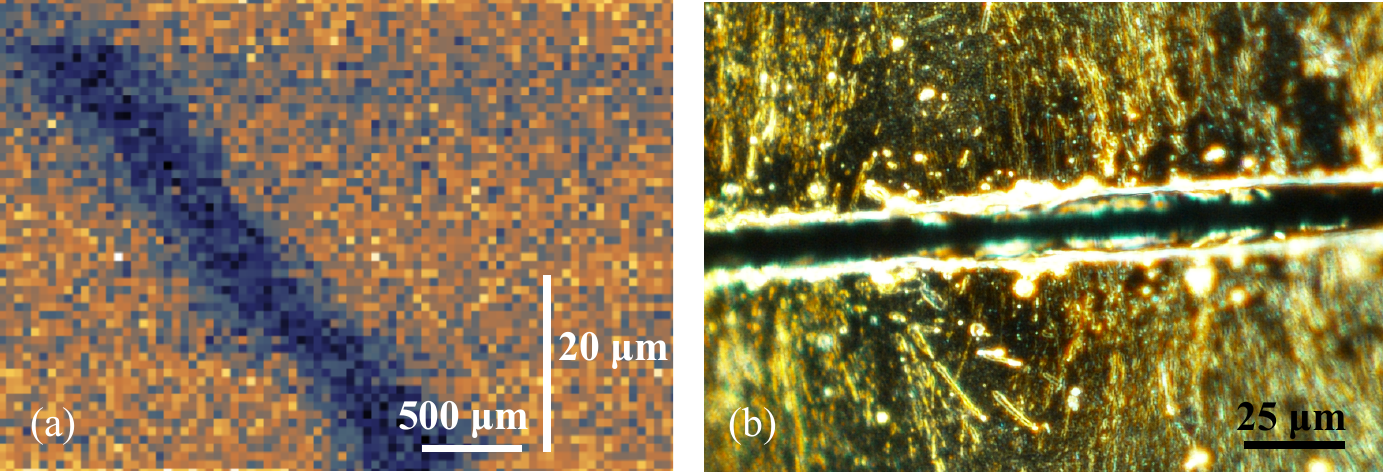}

	\caption{10 $\upmu$m wide slit in a steel foil (a) SHeM image taken with a  4.7 $\upmu$m $\pm $ 0.5 $\upmu$m spot size. The step size in $x$ = 41.2 $\upmu$m and in $y$ = 1 $\upmu$m. A polynomial background has been subtracted from the raw image. (b) Optical microscopy image of a small region of the 10 $\upmu$m slit.} 
	\label{slit1}
\end{figure}

Figure \ref{slit1}(a) presents a SHeM image of a 10 $\upmu$m slit in a stainless steel foil (from National Aperture Inc.). The image was recorded with a spot size (FWHM) of 4.7 $\upmu$m $\pm $ 0.5 $\upmu$m \cite{flatabo2018fast}, a reservoir pressure of 100 bar and a warm source (T $\approx$ 320 K). The step size in $x$ and $y$ are 41.2 $\upmu$m and 1 $\upmu$m, respectively. The imaging time was about 20 hours. 

A slit is arguably not an ideal sample for reflection images as the contrast could simply arise from the transmission of atoms (generating a black/white image). However, the slit is made by laser cutting. This produces sloped edges evident in the SHeM image as a gradual reduction in intensity towards the center of the slit, demonstrating that topography smaller than the lateral resolution of the instrument can be detected. 

\begin{figure}[H]
	\includegraphics[width=0.9\textwidth]{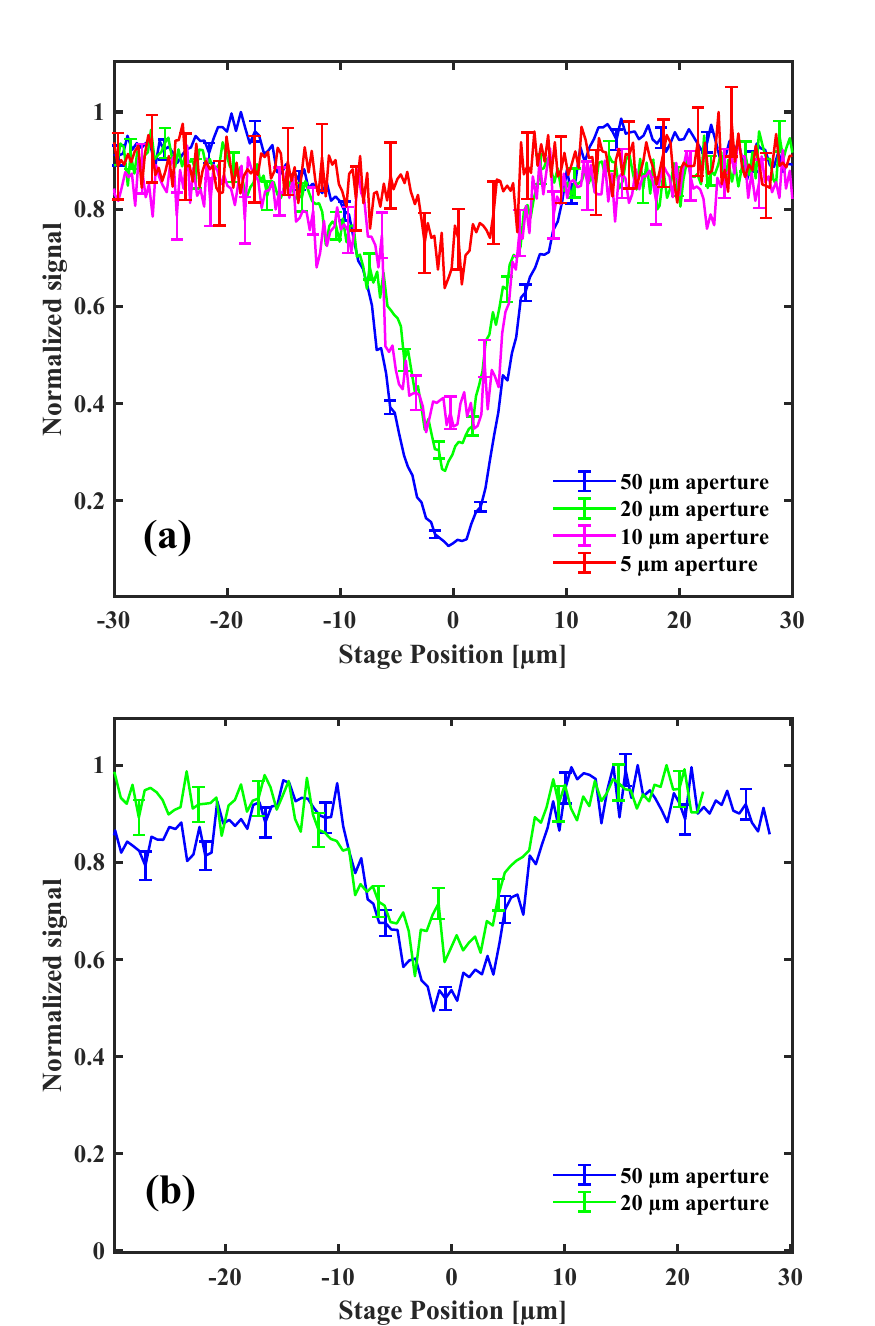}
	\caption{(a) Linescans over a 10 $\upmu$m slit using apertures with a diameter of 50 $\upmu$m (blue), 20 $\upmu$m (green), 10 $\upmu$m (pink) and 5 $\upmu$m (red). The measurements are recorded using a warm source (T $\approx$ 320 K) and a reservoir pressure of 100 bar. (b) Linescans over a 10 $\upmu$m slit using apertures with a diameter of 50 $\upmu$m (blue) and 20 $\upmu$m (green). The measurements are recorded using a cold source (T $\approx$ 140 K) and a reservoir pressure of 100 bar. }
	\label{spotsize}
\end{figure}

In Fig. \ref{spotsize} the beam is scanned over the slit to evaluate the spot size (FWHM) for different collimating apertures. In Fig \ref{spotsize}(a) the source temperature was $\approx$ 320 K  and in Fig \ref{spotsize}(b) the source temperature was $\approx$ 140 K. The reservoir pressure was about 100 bar in both cases. The predicted spot sizes equal the demagnification factor of the instrument (0.226) times the diameter of the respective collimating aperture. In a previous transmission imaging experiment \cite{flatabo2018fast}, the measured spot sizes (FWHM) for aperture diameters down to 10~$\upmu$m were evaluated by fitting the raw data to a double error function, as described by \cite{koch2008imaging}. The model assumes that the slit is a step function over which a Gaussian beam is scanned. This is a suitable model for transmission experiments because the beam cannot penetrate any solid material so imaging in transmission produces a black/white signal. Using this model transmission spot sizes of 10.2 $\upmu$m $\pm$ 0.5 $\upmu$m, 4.7 $\upmu$m $\pm$ 0.5 $\upmu$m and 2.3 $\upmu$m $\pm$ 0.5 $\upmu$m were obtained for a 50 $\upmu$m, 20 $\upmu$m, and  10 $\upmu$m aperture, respectively,  in good agreement with the demagnification factor \cite{flatabo2018fast}. As discussed above in reflection the edges of the slit are detected, and therefore treating the slit as a step function results in a slight overestimation of the diameter of the measured spot size. The spot diameter for the 50 $\upmu$m collimating aperture is fitted to be 11.2 $\upmu$m $\pm$ 1 $\upmu$m rather than 10.2~$\upmu$m measured in transmission, the 20 $\upmu$m collimating aperture is fitted to be 9.5 $\upmu$m $\pm$ 1 $\upmu$m and the spot size of the 10 $\upmu$m collimating aperture is 9 $\upmu$m $\pm$ 1 $\upmu$m. The 5 ~$\upmu$m aperture of the FWHM is fitted to be 2.2 $\upmu$m $\pm$ 1 $\upmu$m, but given the overshoot of the fitting procedure evident in the previous measurement, we are confident that the actual value is close to the predicted spot size of 1.1~$\upmu$m. Note that the decrease in intensity observed in the figures is because in the current instrument configuration, the beams coming from the smaller collimating apertures do not illuminate the whole zone plate~\cite{flatabo2018fast}. This is due to the relatively small diameter of the skimmer currently being used, which selects a slightly too narrow fraction of the expansion for the zone plate to be fully illuminated.  All images in this paper are presented with the spot size that we obtained in transmission.

The experiment was repeated using a cold source, see Fig. \ref{spotsize}(b) with another zone plate designed for a source temperature of 100 K (corresponding to a beam wavelength of 0.098 nm). We only reached a temperature of about 140 K, corresponding to a beam wavelength of 0.083 nm. Therefore the measured spot sizes (FWHM) are slightly bigger: between 14.7 $\upmu$m $\pm$ 1 $\upmu$m and 10.8 $\upmu$m $\pm$ 1 $\upmu$m.

\subsection{Reflection images of an optically transparent resin structure}

\begin{figure}[H]
	\includegraphics[width=\textwidth]{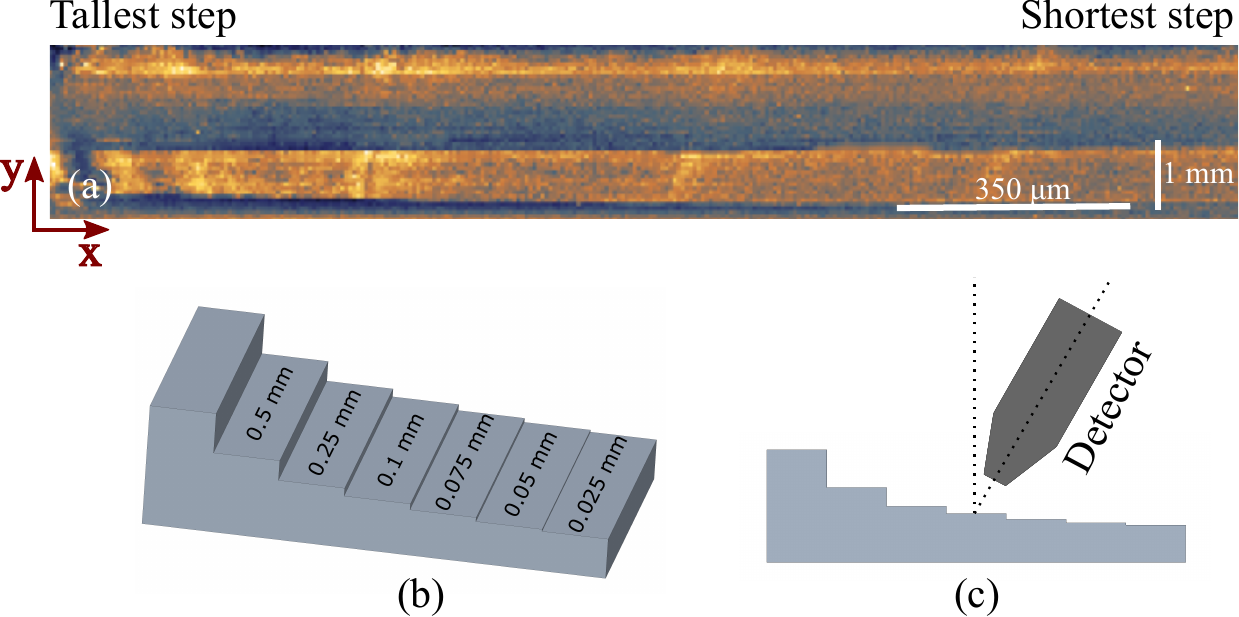}
	\caption{(a) SHeM image of steps in an optically transparent resin. The spot size (FWHM) is  10.2 $\upmu$m $\pm$ 0.5 $\upmu$m.  The step size in $x$ was 10 $\upmu$m and the step size in $y$ was 55 $\upmu$m.  The dark lines in the image are due to thermal drift, see main text. A polynomial background has been subtracted from the raw images. (b) Illustration of the sample including the height of each step in mm. (c) Illustration of the position of the beam and detector relative to the steps.}
	\label{step2}
\end{figure}

Figure \ref{step2} presents a SHeM image of an optically transparent resin (RS-F2-GPCL- 04) "staircase" structure, similar to the sample used by A. Fahy et al. in \cite{fahy2018image}. The sample was imaged using a reservoir pressure of around 110 bar and a warm source (T $\approx$ 320~K)., and a spot size (FWHM) of 10.2 $\upmu$m $\pm$ 0.5 $\upmu$m. The step size in $x$ was 10 $\upmu$m and the step size in $y$ was 55 $\upmu$m, and the imaging time was about 27 hours. A schematic of the resin structure can be seen in Fig. \ref{step2}(b). Fig. \ref{step2}(c) shows the position of the incoming beam relative to the detector. We see that the step's edges are imaged as bright lines. This is because they are rounded, which means that a larger fraction of the signal is scattered in the direction of the detector. The image displays the effect of thermal drift in the form of dark horizontal lines. Temperature fluctuations in the laboratory can be several degrees over a period of hours. This leads to the zone plate position drifting perpendicular to the beam, which again leads to misalignment between the middle-stop and the order-sorting aperture, causing a strong reduction in the beam-spot intensity combined with an increase in background.  

\subsection{Helium atom scattering microscopy: Measuring the spatial scattering distribution from a focused spot} 

\begin{figure}[H]
\includegraphics[width=\textwidth]{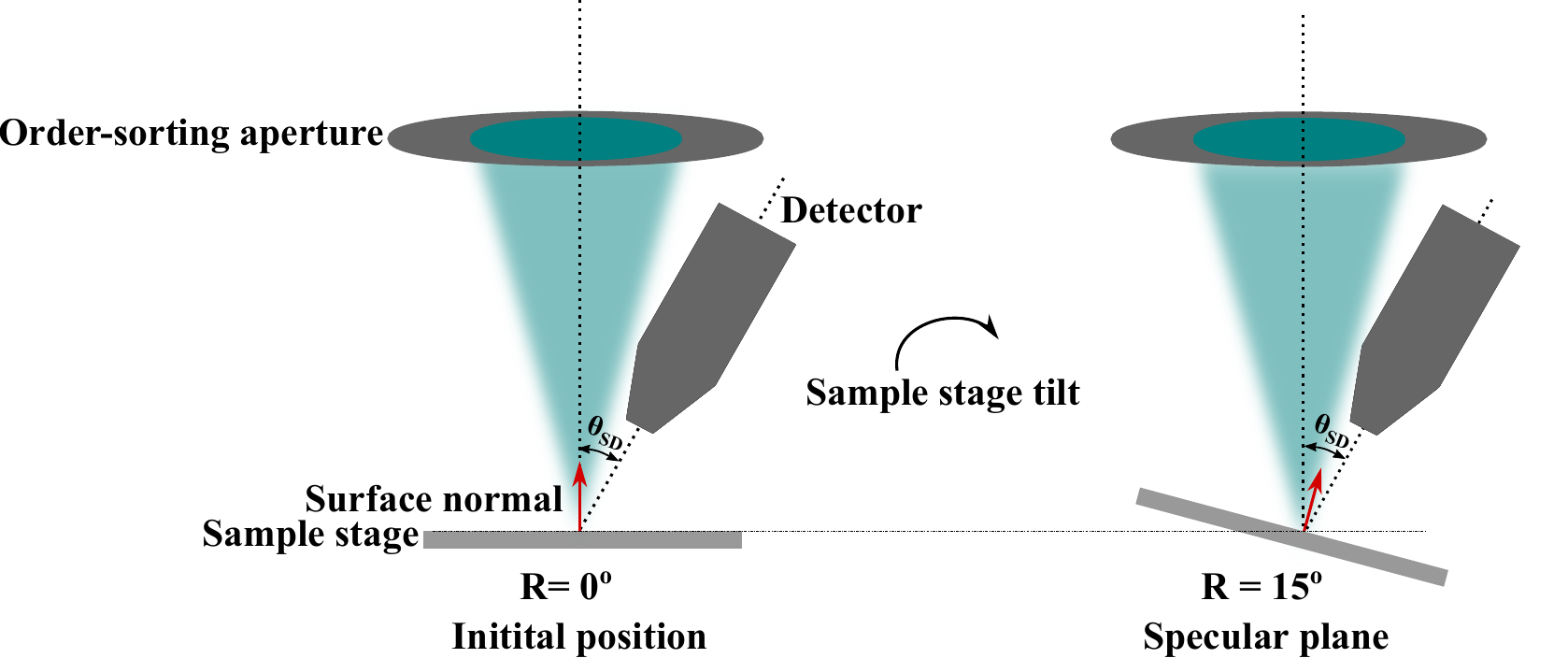}
	\caption{Illustration of the experimental configuration for helium atom scattering microscopy. The angle between the incident beam and the detector $\theta_{SD}$ = 30$^{\circ}$ is fixed. The sample stage is tilted around the specular plane. The movement is corrected so that the beam is kept at a fixed position on the sample during rotation. The specular plane corresponds to $\theta_{f} - \theta_{i} = 0 $ which occurs for a sample stage tilt, R = 15$^{\circ}$.  }
	\label{r1}
\end{figure}

In a standard helium atom scattering (HAS) experiment~\cite{holst2021material} information about the surface structure of a sample is obtained by measuring the spatial distribution of the scattered beam (which is usually a few mm in diameter). For a crystalline sample, this means a diffraction pattern is obtained.  A common configuration is to keep the total scattering angle  $\theta_{SD}$ constant and vary the incident beam angle $\theta_i$ (and thereby also the final beam angle $\theta_f$)  by tilting the sample under the beam. In a recent work~\cite{eder2023sub} SHeM images of macroscopically flat surfaces with roughness on the nano/micron-scale (silicon, gold deposited on silicon, nano- and micro- diamonds and glass and frosted glass) were compared with atomic force microscopy images and HAS scattering data. The HAS measurements revealed that for all surfaces the scattered beam had a broad spatial distribution which looked nearly identical for all the surfaces, with just small variations in broadening. Also, the Time of Flight  (TOF) measurements from all the surfaces were nearly identical in shape. This can be explained by the fact that none of the surfaces had been cleaned in vacuum so the TOF measurements stemmed from physisorbed or chemisorbed contamination (water, CO$_2$, etc.) which was similar for all substrates. Here we are presenting microscopy scale helium atom scattering measurements, tilting the sample stage from  \textit{R} = 0$^{\circ}$ (corresponding to an incoming beam that is perpendicular to the sample stage) to \textit{R} = 30$^{\circ}$, see Fig. \ref{r1}. A sample stage tilt of \textit{R} = 15$^{\circ}$ corresponds to the specular plane.

\begin{figure}[H]
\hspace*{-1in}
\includegraphics[trim={0.6cm 0 0.7cm 0},clip, width=180mm,scale=2]{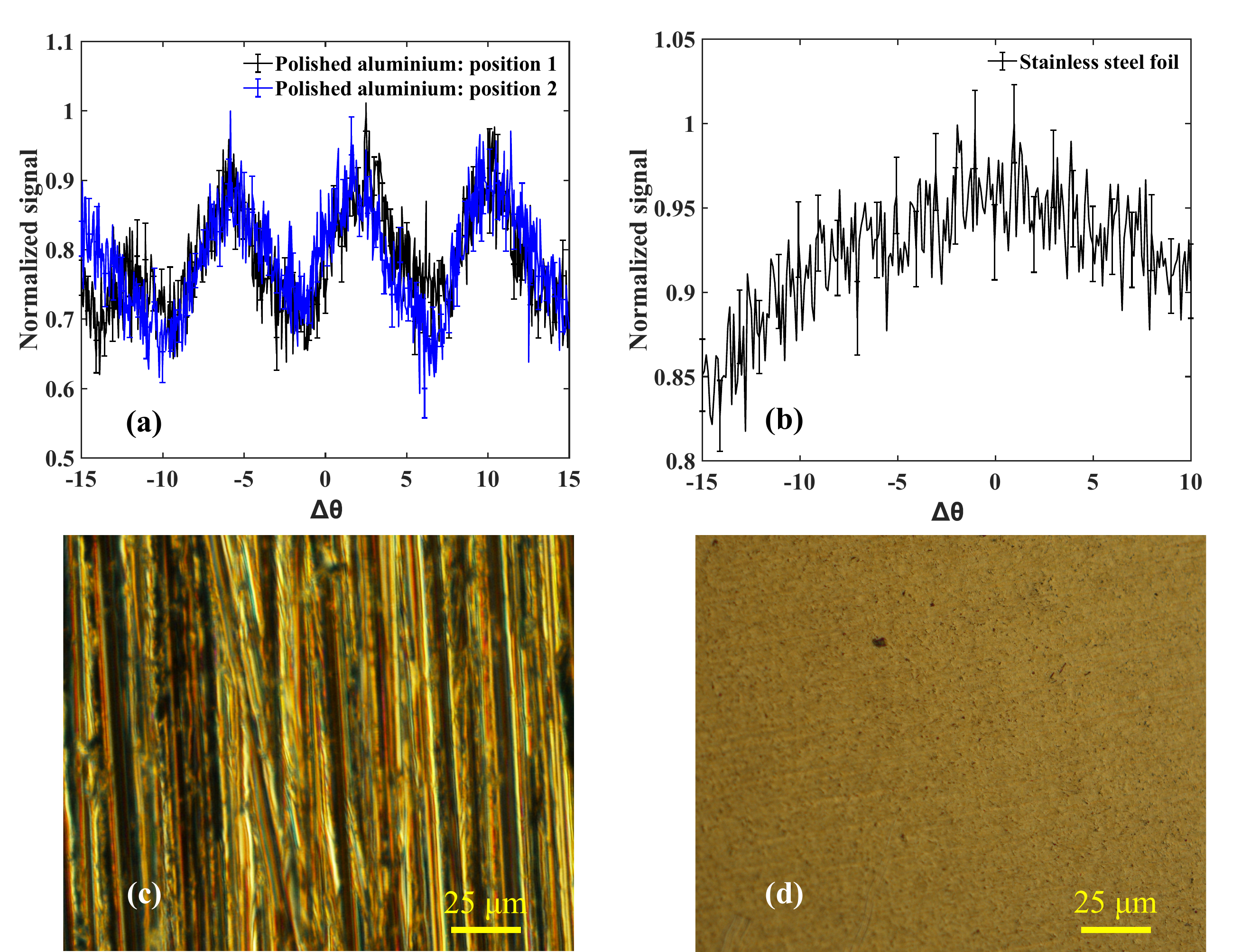}
	\caption{(a) Normalized signal versus $\Delta \theta = \theta_{f}-\theta_{i}$ on two different locations on the aluminium sample holder. The measurements were conducted using a  reservoir pressure of 100 bar, a cold source, and a spot size with a FWHM of about 23 $\upmu$m (b) Normalized signal versus $\Delta \theta = \theta_{f}-\theta_{i}$ for a steel foil. The measurements were conducted using a reservoir pressure of 100 bar, a warm source, and a spot size with a FWHM of  4.7 $\upmu$m $\pm$ 0.5 $\upmu$m (c) Optical microscopy image of the aluminium sample holder (d) Optical microscopy image of the stainless steel foil. }
	\label{r2cold}
\end{figure}

\begin{figure}[H]
\centering
\includegraphics[width=0.7\textwidth]{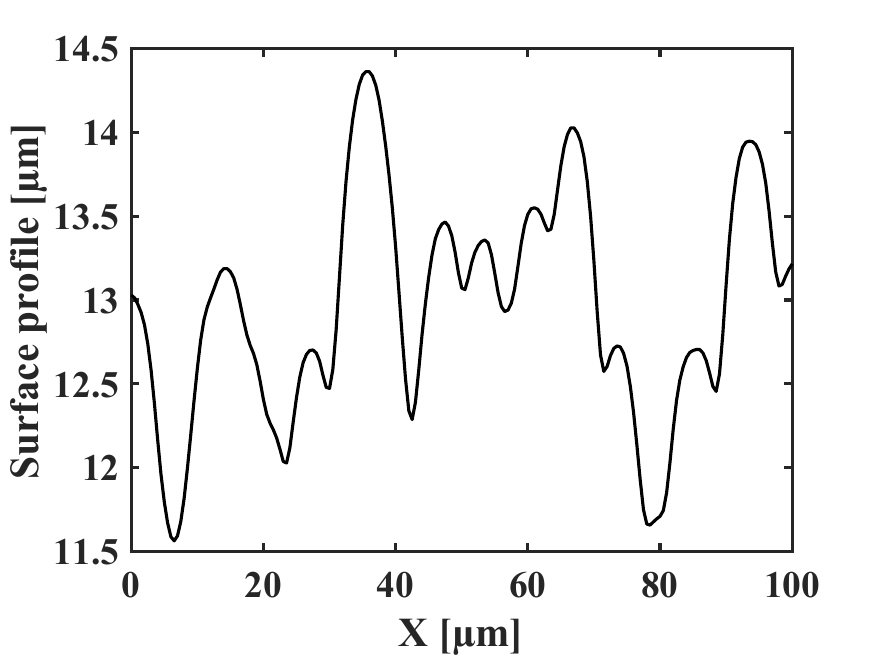}
	\caption{Surface profilometry measurements of the aluminium sample showing the surface topography and revealing a narrow slope range corresponding to what is measured with helium atom scattering microscopy.}
	\label{profil}
\end{figure}

Figure \ref{r2cold}(a) shows the normalized signal versus $\Delta \theta = \theta_f-\theta_i$ for a mechanically polished aluminium surface (the sample holder). Two different locations on the sample holder were measured. The measurements were done using a cold source (T $\approx$ 140 K), a reservoir pressure of 100 bar, and a 100 $\upmu$m aperture (the predicted spot size has a FWHM of about 23 $\upmu$m). The measurements were conducted in one point for all  $\Delta \theta$ with a fixed distance to the detector (see Sec. \ref{exp}). The detected signal shows a strong periodic variation, which we can contribute to the change in scattering direction induced by the narrow slope range of the polishing grooves, see Fig.  \ref{r2cold}(c) for an optical image and Fig. \ref{profil} for a surface profile scan obtained using a profilometer. The surface is oriented so that the polishing grooves are perpendicular to the scattering plane for both the helium microscopy and profilometer measurements. 

Figure \ref{r2cold}(b) shows the normalized signal versus sample tilt for the surface of a stainless steel foil. The measurements are done using a warm source with a FWHM spot size of 4.7 $\upmu$m $\pm$ 0.5 $\upmu$m. An optical microscopy image of the surface is included in Fig. \ref{r2cold}(d). The surface topography differs radically from that of the aluminium sample, and this is reflected in the SHeM measurement. The detected signal increases for increasing $\Delta \theta$  reaching a maximum around the specular plane ($\Delta \theta = 0$) before decaying again. This is very similar to the HAS scattering profiles obtained from nanoscale rough surfaces, with a 3 mm diameter beam, see Fig. 1(a), 2(a) and 3(a) suppl. info in \cite{eder2023sub}, including the slight asymmetry around the specular plane.

\section{Conclusion and Future Work}
In this paper, we present the first reflection SHeM images obtained using a focused helium beam (a helium zone plate microscope). Images are recorded with a spot size (FWHM) down to  4.7 $\upmu$m $\pm $ 0.5 $\upmu$m and focusing down to a spot size of about 1 $\upmu$m is demonstrated. We furthermore present the first helium atom scattering  microscopy experiments measuring the scattering distribution from a focused helium beam spot. As demonstration examples, we show that the scattering distribution of a mechanically polished aluminium surface and a stainless steel foil are radically different, due to the difference in surface roughness. The scattering distribution of the stainless steel foil agrees well with previously macroscopic scale HAS scattering profiles from surfaces with roughness on the nanoscale. 

Future work includes the implementation of a procedure to ensure that the zone plate does not undergo thermal drift. In addition, a bigger diameter skimmer should be used to ensure that the zone plate is fully illuminated for all the collimating apertures. We are confident that these two measures will enable high-quality, sub-micron reflection imaging with the present instrument. Furthermore, experiments are planned which allow the sample to be heated so that surface contamination is removed to enable more information about the true surface structure to be extracted. This is particularly relevant for the HAS microscopy experiments. Long term, the next-generation instruments should be designed according to the model presented in~\cite{palau2017theoretical}. This will ensure the maximum possible intensity for a given resolution and work distance, opening for an instrument with 10~nm resolution. A so-called atom sieve can be used to overcome the challenges related to the requirement of a 10~nm outermost zone~\cite{eder2015, Flataboe2017}.

\section{Acknowledgements}
The authors thank Paul C. Dastoor and the  SHeM group at the Centre for Organic Electronics, University of Newcastle, Australia for the 3D-printed staircase structure which was produced by Sabrina D. Eder, Matt Barr and Adam Fahy. We further thank Marit Hougen and Paul Johan Høl for help with the profilometer measurements and Henry I Smith at Massachusetts Institute of Technology for support in fabricating the zone plates. Ranveig Flatabø acknowledges funding from the University of Bergen Faculty for Mathematics and Natural Sciences. The development of the zone plate helium atom microscope has been supported by the Trond Mohn Research Foundation and the EU project NEMI, FP7-NMP-2012-SME-6, Grant number 309672. 

\newpage

\bibliographystyle{unsrt}
\bibliography{bib}

\end{document}